\pdfoutput=1
\documentclass[runningheads,10pt]{llncs}

\usepackage[T1]{fontenc}
\usepackage{amsmath,amssymb}
\usepackage[hidelinks]{hyperref}
\usepackage[sectionbib]{natbib}
\usepackage{doi}
\usepackage{xcolor}
\usepackage{graphicx}
\usepackage{setspace}
\usepackage{enumitem}
\usepackage[a4paper,
            bindingoffset=0.0cm,
            left=2.54cm,
            right=2.54cm,
            top=2.54cm,
            bottom=2.54cm,
            footskip=0.0cm]{geometry}
\usepackage{booktabs}
\usepackage{upgreek}
\usepackage{algorithm}
\usepackage{algpseudocode}
\usepackage{float}
\usepackage{soul}

\IfFileExists{orcid.pdf}{}{}

\begin{document}

\title{A Two-Step Ensemble Score Filter for Data Assimilation\\ in Partially Observed Systems}

\author{Zixiang Xiong\inst{1}\thanks{This preprint has not been peer-reviewed and is shared by the authors to support the timely, noncommercial dissemination of research findings. Copyright remains with the authors, and the manuscript may not be copied or reposted without their explicit permission.}
\and Feng Bao\inst{1}
\and Hristo G.~Chipilski\inst{2}
\and Siming Liang\inst{3}
\and Jingqiao Tang\inst{1}
\and Guannan Zhang\inst{3}}
\authorrunning{Xiong et al.}

\institute{Department of Mathematics, Florida State University, Tallahassee, FL
\and Department of Scientific Computing, Florida State University, Tallahassee, FL
\and Computer Science and Mathematics Division, Oak Ridge National Laboratory, Oak Ridge, TN\\
Corresponding author: \email{hchipilski@fsu.edu}}

\maketitle

\vspace*{-4mm}
\begin{abstract}
Data assimilation blends model forecasts with observations to estimate the evolving state of complex dynamical systems, but sparse observing networks remain challenging because unobserved state variables are not directly constrained by observations. In this work, we introduce the Ensemble Score Filter with Linear Regression (EnSF-LR), a two-step filtering method for partially observed nonlinear systems. At each analysis time, EnSF-LR first applies the Ensemble Score Filter (EnSF) to update the observed state components using a nonlinear score-based analysis update. It then computes the resulting observed-state analysis increments and maps these corrections to the unobserved components through the ensemble-based prior covariance matrix. The latter amounts to the same linear regression mechanism used by Ensemble Kalman Filters (EnKFs). We evaluate EnSF-LR using the Lorenz-63 and 40-dimensional Lorenz-96 systems with sparse linear and nonlinear observations. The method is compared with the original EnSF and with the classical stochastic EnKF. In the linear-observation experiments, EnSF-LR produces accuracy comparable to the EnKF baseline while substantially reducing error relative to the original EnSF. In the nonlinear-observation experiments, EnSF-LR achieves lower full-state root-mean-square error than both the original EnSF and the EnKF reference. These results suggest that hybridizing score-based and EnKF analysis schemes provides an effective strategy for assimilating sparse and nonlinear observations.

\keywords{data assimilation \and Ensemble Score Filter \and Ensemble Kalman Filter \and partially observed systems \and nonlinear observations}
\end{abstract}

\subsubsection{Significance statement}

In many prediction problems, such as weather and climate applications, only part of the system can be directly observed, but scientists still need to estimate the full system state. This study develops a new ensemble data assimilation method that first corrects the observed variables using generative AI (GenAI) techniques and then uses statistical relationships within the ensemble to adjust the unobserved variables. Experiments with two standard chaotic models show that the proposed method improves full-state estimation compared with standard approaches and previously developed GenAI-based data assimilation methods when observations are sparse and nonlinear.


\section{Introduction}
\label{section:introduction}

Data assimilation (DA) combines numerical model forecasts with observations to estimate the evolving state of geophysical systems. This problem is central to numerical weather prediction, ocean forecasting, and climate applications, where the true state is high-dimensional, nonlinear, chaotic, and only partially observed \citep{kalnay2003atmospheric,evensen2009data,asch2016data,reich2015probabilistic}. Because observations are often sparse in space and time, an effective DA method must not only correct directly observed variables but also project observational information to unobserved components of the state.

Since the work of \citet{evensen1994sequential}, ensemble Kalman filtering has become a central approach for sequential data assimilation. In these methods, an ensemble of forecasts is used not only to represent uncertainty, but also to estimate flow-dependent covariance relationships among state variables \citep{houtekamer1998data,burgers1998analysis,whitaker2002ensemble}. These ensemble-estimated covariances are crucial in partial-observation settings because they provide a mechanism for spreading information from observed variables to unobserved parts of the state. In this way, EnKF-type methods can update variables that are not directly measured, provided that they are statistically correlated with the observed quantities in the forecast ensemble.

A useful interpretation of this information-transfer mechanism was provided by \citet{anderson2003local}, who formulated ensemble filtering in a joint state--observation space. In this view, the assimilation update can be organized into two steps. First, the ensemble estimate of the observed quantity is updated. Second, the resulting observation-space increments are mapped back to the model state through linear regression. This two-step perspective clarifies how EnKF methods propagate observational information from observed to unobserved components through ensemble-estimated cross-covariances. It also provides a natural framework for separating the problem of correcting observed quantities from the problem of spreading those corrections across the full model state.

Despite their success, standard EnKF methods are based on linear analysis updates derived from Gaussian assumptions. These assumptions can be restrictive when the prior distribution is strongly non-Gaussian or when the observation operator is nonlinear \citep{vanleeuwen2009particle,law2015data,reich2015probabilistic}. Nonlinear observation operators can produce asymmetric or non-Gaussian likelihoods, and the resulting analysis increments may not be well represented by a linear update. These difficulties are amplified in chaotic systems with sparse observations, where forecast errors can grow rapidly between assimilation times and where unobserved variables must be inferred indirectly.

To address departures from Gaussian behavior, a number of non-Gaussian data assimilation methods have been developed. Particle filters provide a direct Bayesian approach by representing the posterior distribution with weighted ensemble samples rather than assuming a Gaussian analysis distribution \citep{gordon1993novel,crisan2002survey,vanleeuwen2009particle,vanleeuwen2019particle}. In high-dimensional geophysical applications, however, standard particle filters often suffer from weight degeneracy, in which only a small number of particles receive appreciable posterior weight \citep{SnyderEtAl2008, vanleeuwen2019particle}. This limitation can make their direct application impractical unless very large ensembles are used. As a result, several alternative ensemble-based methods have been proposed to improve non-Gaussian filtering performance, including particle flow filters \citep{pulido2019sequential,hu2021particle}, lognormal and bi-Gaussian extensions of existing data assimilation methods \citep{fletcher2010mixed,chan2020bi,fletcher2023data}, and quantile-conserving ensemble filters \citep{anderson2022quantile,anderson2023quantile}.

Recent advances at the intersection of DA and AI have introduced additional tools for nonlinear and non-Gaussian filtering. Machine learning approaches have been used to learn model dynamics from sparse observations, estimate model error and parameters, perform DA in lower-dimensional latent spaces, and couple DA methods with data-driven forecast models \citep{bocquet2019data,bocquet2020bayesian,brajard2020combining,bonavita2020model,farchi2021online,peyron2021latent,chattopadhyay2023data,meng2024assimilating}. In particular, generative modeling provides a promising direction for non-Gaussian DA because generative models are designed to represent and sample from complex probability distributions \citep{song2021score,luo2022understanding,chipilski2025conjugate}. This connection has recently motivated the development of diffusion model-based ensemble DA methods \citep{rozet_louppe_2023,manshausen_et_al_2024,martin2025generative}, including the Ensemble Score Filter (EnSF) that this work focuses on \citep{bao2024score,bao2025nonlinear,liang2025inpainting,xiong2026robustness,hu2026twophase,zhang2025iensf}.

Unlike many ML-based diffusion approaches, EnSF is a training-free approach: the prior score is approximated directly from the forecast ensemble, avoiding the need to train a neural score model at every assimilation cycle. This feature makes EnSF attractive for nonlinear DA problems in which non-Gaussian analysis corrections are important. Recent work has demonstrated the potential of EnSF in high-dimensional Lorenz-96 systems and in a surface quasigeostrophic (SQG) model, where it shows advantages over Kalman-type ensemble methods in nonlinear-observation and imperfect-model settings \citep{bao2025nonlinear}. Follow-up studies have further investigated EnSF robustness to observation-network design, partial-observation settings with image inpainting, porous-media flow assimilation, real-data forecast correction, and iterative posterior-score refinement \citep{xiong2026robustness,liang2025inpainting,hu2026twophase,hu2026energy,zhang2025iensf}. These results suggest that score-based ensemble filtering provides a promising framework for nonlinear and non-Gaussian DA.

However, sparse observations introduce a fundamental difficulty for the original EnSF. When the observation operator depends only on a subset of the state variables, the likelihood-gradient term in the score-based analysis update directly acts only on the observed components. As shown later in Eq.~\eqref{eq:likelihood_gradient_partition}, the likelihood-gradient correction is nonzero only for the observed components; the unobserved components are affected only indirectly through the prior score and subsequent model dynamics. As a result, EnSF accurately corrects observed variables while leaving larger errors in unobserved components, particularly under sparse observing networks.

Motivated by \citet{anderson2003local}'s two-step ensemble filtering framework, we introduce the Ensemble Score Filter with Linear Regression (EnSF-LR), a two-step method for DA under sparse observations. At each analysis time, EnSF-LR first applies the score-based EnSF update to the observed components. It then maps the resulting analysis increments to the unobserved components through a linear regression step. In this way, the method hybridizes the nonlinear analysis capability of EnSF in observation space with the covariance-based regression mechanism underlying EnKF algorithms.

The remainder of the paper is organized as follows.
Section~\ref{section:methodology} introduces the Bayesian filtering framework, reviews the EnSF, presents the proposed EnSF-LR method, and describes the EnKF reference methods used for comparison.
Section~\ref{section:experimental_design} describes the model dynamics, observation configurations, evaluation metrics, and ensemble settings.
Section~\ref{section:results} presents the numerical results for the linear and nonlinear observation experiments.
Section~\ref{section:conclusion} summarizes the main findings and discusses directions for future work.

\section{Methodology}
\label{section:methodology}
\subsection{Bayesian Framework for Data Assimilation}

In the data assimilation problem, we consider the following state-space model:
\begin{align}
    \text{State:} \quad
    \mathbf{x}_t &= \mathcal{F}(\mathbf{x}_{t-1}, \boldsymbol{\omega}_t),
    \label{eq:model} \\
    \text{Observation:} \quad
    \mathbf{y}_t &= \mathcal{H}(\mathbf{x}_t) + \boldsymbol{\epsilon}_t,
    \label{eq:obs}
\end{align}
where $\mathbf{x}_t \in \mathbb{R}^d$ denotes the model state at time $t$, which is the primary quantity of interest in this work, and $\mathbf{y}_t \in \mathbb{R}^p$ denotes the corresponding observation. Here, $\mathcal{F}$ represents the nonlinear forecast model, while $\mathcal{H}$ denotes the observation operator, which may be either linear or nonlinear. The goal of data assimilation is to estimate the filtering distribution $p(\mathbf{x}_t \mid \mathbf{y}_{1:t})$, where $\mathbf{y}_{1:t}:={\mathbf{y}_1,\ldots,\mathbf{y}_t}$ denotes the collection of observations up to time $t$. The model error $\boldsymbol{\omega}_t$ and observation error $\boldsymbol{\epsilon}_t$ are assumed to be independent zero-mean Gaussian random variables with covariance matrices $\mathbf{Q}_t$ and $\mathbf{R}_t$, respectively.

We formulate data assimilation as a sequential Bayesian inference problem. Sequential Bayesian filtering consists of alternating prediction and analysis steps. In the prediction step, the posterior distribution at time $t-1$ is propagated forward through the forecast model to obtain the prior distribution at time $t$:
\begin{align}
    p(\mathbf{x}_t \mid \mathbf{y}_{1:t-1})
    =
    \int
    p(\mathbf{x}_t \mid \mathbf{x}_{t-1})
    p(\mathbf{x}_{t-1} \mid \mathbf{y}_{1:t-1})
    \, d\mathbf{x}_{t-1},
\end{align}
where $p(\mathbf{x}_t \mid \mathbf{x}_{t-1})$ is determined by the forecast model in Eq.~\eqref{eq:model}.

In the analysis step, the prior distribution is updated using the new observation $\mathbf{y}_t$ through Bayes' theorem:
\begin{align}
    p(\mathbf{x}_t \mid \mathbf{y}_{1:t})
    =
    \frac{
    p(\mathbf{y}_t \mid \mathbf{x}_t)
    p(\mathbf{x}_t \mid \mathbf{y}_{1:t-1})
    }{
    p(\mathbf{y}_t \mid \mathbf{y}_{1:t-1})
    },
\end{align}
where $p(\mathbf{y}_t \mid \mathbf{y}_{1:t-1})$ is a normalizing constant independent of $\mathbf{x}_t$. Thus, $ p(\mathbf{x}_t \mid \mathbf{y}_{1:t})
    \propto
    p(\mathbf{y}_t \mid \mathbf{x}_t)
    p(\mathbf{x}_t \mid \mathbf{y}_{1:t-1})$.
Under the observation model in Eq.~\eqref{eq:obs} with additive Gaussian
observation error $\boldsymbol{\epsilon}_t \sim \mathcal{N}(\mathbf{0},\mathbf{R})$,
the likelihood is given by
\begin{equation}
\begin{aligned}
p(\mathbf{y}_t \mid \mathbf{x}_t)
&=
\frac{1}{\sqrt{(2\pi)^p \det(\mathbf{R})}}
\exp \left[
-\frac{1}{2}
\left(
\mathbf{y}_t - \mathcal{H}(\mathbf{x}_t)
\right)^\top
\mathbf{R}^{-1}
\left(
\mathbf{y}_t - \mathcal{H}(\mathbf{x}_t)
\right)
\right].
\end{aligned}
\label{eq:likelihood}
\end{equation}

\subsection{Overview of Ensemble Score Filter}
\label{subsection:2.b}

The Ensemble Score Filter (EnSF) is a diffusion-based ensemble DA method that samples the filtering density $p(\mathbf{x}_t \mid \mathbf{y}_{1:t})$ through a score-based generative process. Rather than training a complex neural network, EnSF estimates the prior score directly from the forecast ensemble and then incorporates the new observations through an approximate posterior score. In this framework, samples are transported through a pseudo-time diffusion process between the filtering distribution and a reference Gaussian distribution, providing a flexible nonlinear alternative to Gaussian-based EnKF approaches.

Here, we follow the EnSF formulation described by \citet{bao2024ensemble}, including its training-free ensemble score approximation and the use of a damped likelihood-gradient correction during the analysis step. We therefore omit the detailed SDE derivation and score-estimation formulas, and focus instead on how the EnSF update is modified for the partial observation setting considered in this study.

To clarify the behavior of EnSF under sparse observations, we decompose the state vector into observed and unobserved components:
\begin{align}
    \mathbf{x}_t
    =
    \begin{bmatrix}
    \mathbf{x}^{obs}_t \\
    \mathbf{x}^{unobs}_t
    \end{bmatrix},
    \qquad
    \mathbf{x}^{obs}_t\in\mathbb{R}^p,
    \qquad
    \mathbf{x}^{unobs}_t\in\mathbb{R}^{d-p}.
    \label{eq:state_partition}
\end{align}
This decomposition defines the observed component as the part of the state on which the observation operator acts; hence,
\begin{align}
    \mathcal{H}(\mathbf{x}_t)
    =
    \mathcal{H}(\mathbf{x}^{obs}_t).
\end{align}
Then the likelihood-gradient term can be written as
\begin{align}
\label{eq:likelihood_gradient_partition}
    \nabla_{\mathbf{x}}
    \log p(\mathbf{y}_t\mid \mathbf{x}_t)
    =
    \begin{bmatrix}
    \left(
    \dfrac{\partial \mathcal{H}}{\partial \mathbf{x}^{obs}_t}
    \right)^{\top}
    \mathbf{R}^{-1}
    \left[
    \mathbf{y}_t
    -
    \mathcal{H}(\mathbf{x}^{obs}_t)
    \right]
    \\
    \mathbf{0}
    \end{bmatrix}.
\end{align}
Equation~\eqref{eq:likelihood_gradient_partition} shows that the likelihood-gradient correction is nonzero only in the observation subspace. This occurs because, under the sparse observation setting considered here, the observation operator depends only on $\mathbf{x}^{obs}_t$ and has no explicit dependence on $\mathbf{x}^{unobs}_t$. Consequently, the derivative of the log-likelihood with respect to the unobserved components is zero.

This does not imply that the unobserved variables are dynamically unrelated to the observed variables. Rather, it means that the likelihood term used in the standard EnSF analysis update does not directly encode this cross-variable dependence. In the EnSF formulation considered here, the prior distribution is represented empirically by the forecast ensemble, without an explicit covariance matrix that maps the likelihood correction from observed to unobserved variables. Therefore, the original EnSF analysis step does not directly modify \(\mathbf{x}^{unobs}\); these components are affected only later through the forecast dynamics. Under sparse observations, this can lead to accurate analysis of the observed variables while leaving substantial errors in the unobserved variables. This observation motivates the regression-based extension introduced in Section~\ref{subsection:2.c}.

\subsection{EnSF via Linear Regression}
\label{subsection:2.c}

As discussed in Eq.~\eqref{eq:likelihood_gradient_partition}, under sparse observations the likelihood-gradient term in the standard EnSF update directly acts only on the observed components of the state vector. Consequently, the unobserved components may remain insufficiently corrected during the analysis step. To address this limitation, we augment EnSF with a linear regression update that maps the observation-space analysis correction to the unobserved components through the prior covariances. The resulting method, referred to as EnSF-LR, consists of two steps.

\begin{itemize}
    \item \textbf{Step 1 (Observed State Update):}
    Given the prior ensemble $\{\mathbf{x}^{(i)}_{t|t-1}\}_{i=1}^{N_e}$, we partition each ensemble member into observed and unobserved components,
    \begin{equation}
        \mathbf{x}^{(i)}_{t|t-1}
        =
        \left(
        \mathbf{x}^{obs,(i)}_{t|t-1},
        \mathbf{x}^{unobs,(i)}_{t|t-1}
        \right),
        \qquad i=1,\ldots,N_e .
        \label{eq:ensflr_partition}
    \end{equation}
    We then apply EnSF to the observed components to assimilate the new observation $\mathbf{y}_t$, producing the updated observed-state ensemble
    $\{\mathbf{x}^{obs,(i)}_{t|t}\}_{i=1}^{N_e}$. At this stage, the unobserved components are not directly modified by the score-based likelihood update and remain at their prior values,
    $\{\mathbf{x}^{unobs,(i)}_{t|t-1}\}_{i=1}^{N_e}$.

    The observed-state analysis increment for each ensemble member is defined as
    \begin{equation}
        \Delta \mathbf{x}^{obs,(i)}_t
        =
        \mathbf{x}^{obs,(i)}_{t|t}
        -
        \mathbf{x}^{obs,(i)}_{t|t-1},
        \qquad i=1,\ldots,N_e .
        \label{eq:obs_increment_step1}
    \end{equation}

    \item \textbf{Step 2 (Unobserved State Update):}
    To update the unobserved components, we estimate the statistical relationship between observed and unobserved variables from the prior ensemble. Specifically, we compute the observed-state covariance
    \begin{equation}
        \mathbf{P}_{oo}
        =
        \mathrm{Cov}
        \left(
        \mathbf{x}^{obs}_{t|t-1},
        \mathbf{x}^{obs}_{t|t-1}
        \right),
        \label{eq:Poo}
    \end{equation}
    and the cross-covariance between the unobserved and observed components,
    \begin{equation}
        \mathbf{P}_{uo}
        =
        \mathrm{Cov}
        \left(
        \mathbf{x}^{unobs}_{t|t-1},
        \mathbf{x}^{obs}_{t|t-1}
        \right),
        \label{eq:Puo}
    \end{equation}
    where both covariance matrices are estimated from the prior ensemble at the current assimilation time. The unobserved-state increment for each ensemble member is then obtained through the linear regression relation
    \begin{equation}
        \Delta \mathbf{x}^{unobs,(i)}_t
        =
        \mathbf{P}_{uo}
        \mathbf{P}_{oo}^{-1}
        \Delta \mathbf{x}^{obs,(i)}_t,
        \qquad i=1,\ldots,N_e .
        \label{eq:unobs_increment}
    \end{equation}

In the experiments below, the action of $\mathbf{P}_{oo}^{-1}$ is computed by solving the corresponding linear system rather than explicitly forming the inverse. Then the posterior unobserved ensemble is updated as
\begin{equation}
    \mathbf{x}^{unobs,(i)}_{t|t}
    =
    \mathbf{x}^{unobs,(i)}_{t|t-1}
    +
    \Delta \mathbf{x}^{unobs,(i)}_t,
    \qquad i=1,\ldots,N_e .
    \label{eq:unobs_update}
\end{equation}

Thus, the ensemble-estimated cross-covariance structure projects the EnSF analysis correction from the observation space to the unobserved state variables.
\end{itemize}

Finally, the updated observation-space ensemble $\{\mathbf{x}^{obs,(i)}_{t|t}\}_{i=1}^{N_e}$ from Step 1 is combined with the updated unobserved ensemble $\{\mathbf{x}^{unobs,(i)}_{t|t}\}_{i=1}^{N_e}$ from Step 2 to form the complete posterior ensemble,
\begin{equation}
    \mathbf{x}^{(i)}_{t|t}
    =
    \left(
    \mathbf{x}^{obs,(i)}_{t|t},
    \mathbf{x}^{unobs,(i)}_{t|t}
    \right),
    \qquad i=1,\ldots,N_e .
    \label{eq:ensflr_full_posterior}
\end{equation}
This completes the analysis step at time $t$. The resulting posterior ensemble is then propagated forward by the forecast model to begin the next prediction--analysis cycle.

\begin{algorithm}[H]
\caption{EnSF-LR Data Assimilation}
\label{alg:ensf-lr}
\begin{algorithmic}[1]
\State \textbf{Input:} Posterior ensemble $\{\mathbf{x}^{(i)}_{t-1|t-1}\}_{i=1}^{N_e}$, observation $\mathbf{y}_t$
\State \textbf{Forecast:} Propagate ensemble through $\mathcal{F}$ to obtain $\{\mathbf{x}^{(i)}_{t|t-1}\}_{i=1}^{N_e}$ and partition into $(\mathbf{x}^{obs}_{t|t-1},\mathbf{x}^{unobs}_{t|t-1})$
\State \textbf{Observed update:} Apply EnSF to the observed components and compute $\Delta \mathbf{x}^{obs}_t=\mathbf{x}^{obs}_{t|t}-\mathbf{x}^{obs}_{t|t-1}$
\State \textbf{Covariance estimation:} Estimate $\mathbf{P}_{oo}=\mathrm{Cov}(\mathbf{x}^{obs}_{t|t-1},\mathbf{x}^{obs}_{t|t-1})$ and $\mathbf{P}_{uo}=\mathrm{Cov}(\mathbf{x}^{unobs}_{t|t-1},\mathbf{x}^{obs}_{t|t-1})$ from the prior ensemble
\State \textbf{Regression update:} Compute $\Delta \mathbf{x}^{unobs}_t=\mathbf{P}_{uo}\mathbf{P}_{oo}^{-1}\Delta \mathbf{x}^{obs}_t$
\State \textbf{Combine:} Set $\mathbf{x}^{unobs}_{t|t}=\mathbf{x}^{unobs}_{t|t-1}+\Delta \mathbf{x}^{unobs}_t$ and form $\mathbf{x}_{t|t}=(\mathbf{x}^{obs}_{t|t},\mathbf{x}^{unobs}_{t|t})$
\State \textbf{Output:} Posterior ensemble $\{\mathbf{x}^{(i)}_{t|t}\}_{i=1}^{N_e}$
\end{algorithmic}
\end{algorithm}

\subsection{Other closely related two-step DA methods}
\label{sec:other_2step_methods}

The regression step in EnSF-LR is motivated by the broader two-step ensemble filtering framework of \citet{anderson2003local}, the Bayesian interpretation of two-step ensemble filters developed by \citet{grooms2022comparison}, and the recent high-dimensional implementation of this idea in the particle flow filtering algorithm of \citet{hu2024pff}. In a standard two-step ensemble filter, the first step updates the ensemble in a reduced space associated with the observations, and the second step maps the resulting increments back to the model state through covariance-based regression. In the terminology of \citet{hu2024pff}, this reduced space is closely related to the inner domain used for the nonlinear update. In the present paper, the role of this reduced space is played by $\mathbf{x}^{obs}$, defined as the subset of state variables that enter the observation operator $\mathcal{H}(\cdot)$.

There is, however, an important distinction between EnSF-LR and the corresponding two-step particle flow filter of \citet{hu2021particle}. EnSF-LR is not formulated as a serial algorithm that assimilates observations one at a time. Instead, the selected observed state variables are updated jointly by EnSF, and the resulting observed-state increments are then mapped to the unobserved variables through the regression operator $\mathbf{P}_{uo}\mathbf{P}_{oo}^{-1}$. This joint update is deliberate: it allows the present study to isolate the approximation error of the proposed two-step filtering update, without introducing additional practical choices associated with serial processing, such as localization, inflation, or assumptions about the sequential treatment of observations. Developing a serial version of EnSF-LR that can be implemented in popular software packages like the Data Assimilation Research Testbed \citep[DART; ][]{elgharamti2025dart} is left for future work.

\subsection{EnKF benchmark}
\label{subsection:2.d}

The Ensemble Kalman Filter (EnKF) is used as the covariance-based ensemble DA baseline in this study
\citep{evensen1994sequential,houtekamer1998data,burgers1998analysis}. As a Monte Carlo approximation of the
Kalman filter, the stochastic EnKF estimates forecast-error covariances from an ensemble of state vectors
and therefore avoids explicitly evolving tangent linear models.

We consider the standard stochastic EnKF update and its equivalent two-step regression formulation.
In the standard formulation, the full state vector is updated directly using the Kalman gain estimated
from the forecast ensemble. In the two-step formulation, following the joint state--observation-space
interpretation of \citet{anderson2003local}, the ensemble is first updated in observation space, and the
resulting observation-space increments are mapped back to the model state through an ensemble-estimated
linear regression relation.

Under the covariance definitions used in this study, these two formulations produce the same analysis
update. Therefore, throughout the main numerical results we report a single EnKF baseline rather than
treating the one-step and two-step implementations as distinct methods. The equivalence is shown
mathematically in Appendix~\ref{app:enkf_equivalence}, and a numerical validation is provided in
Fig.~\ref{fig:enkf_equivalence}.

\begin{table}[ht!]
\caption{Summary of filtering methods compared in this study.}
\label{tab:methods}
\centering
\small
\begin{tabular}{p{0.18\textwidth}p{0.72\textwidth}}
\toprule
Method & Description \\
\midrule
EnSF &
Original Ensemble Score Filter; see Section~\ref{subsection:2.b}. \\

EnSF-LR &
EnSF observed-state update combined with a regression-based update for
unobserved components; see Section~\ref{subsection:2.c}. \\

EnKF &
Stochastic EnKF baseline. The equivalence between the one-step and
two-step formulations is discussed in Appendix~\ref{app:enkf_equivalence}. \\
\bottomrule
\end{tabular}
\end{table}

\section{Experimental Design}
\label{section:experimental_design}
\subsection{Forecast models}

We evaluate the performance of the EnSF-LR by comparing it with the EnKF on two benchmark chaotic systems: the Lorenz-63 and Lorenz-96 models. These models are widely used in data assimilation studies because they provide highly nonlinear dynamical settings for assessing filter accuracy, stability, and robustness under partial observations. The Lorenz-63 model is used as a low-dimensional test case in which the behavior of the filters can be visualized directly, while the Lorenz-96 model provides a higher-dimensional cyclic system with spatially propagating errors and more complex cross-variable dependencies. Together, these two systems allow us to examine how the proposed regression-based update strategy in EnSF improves state estimation performance across idealized chaotic dynamical systems. In our numerical experiments, we consider the perfect-model setting and do not add stochastic model error during the forecast step.

\noindent $\bullet$ \textbf{Lorenz-63:}
    The Lorenz-63 system is a three-dimensional chaotic model governed by
    \begin{align}
        \frac{dx}{dt} &= \sigma(y-x), \label{eq:lorenz63_x}\\
        \frac{dy}{dt} &= x(\rho-z)-y, \label{eq:lorenz63_y}\\
        \frac{dz}{dt} &= xy-\beta z, \label{eq:lorenz63_z}
    \end{align}
    with the standard chaotic parameters $\sigma=10$, $\rho=28$, and $\beta=8/3$. In this configuration, the system exhibits the familiar two-lobed chaotic attractor and strong nonlinear coupling among the three state variables. These properties make Lorenz-63 useful for diagnosing how observational information from a single observed variable is mapped to unobserved variables. In all experiments, the Lorenz-63 system is integrated using \texttt{scipy.integrate.odeint}, which calls the LSODA solver from the
    ODEPACK library and automatically switches between nonstiff Adams and stiff
    BDF methods \citep{virtanen2020scipy,hindmarsh1983odepack,petzold1983automatic}.

\noindent $\bullet$ \textbf{Lorenz-96:}
    The Lorenz-96 system is a forced dissipative model commonly used as an intermediate-dimensional testbed for atmospheric data assimilation. It is defined by
    \begin{align}
        \frac{dx_i}{dt}
        =
        (x_{i+1}-x_{i-2})x_{i-1}
        -
        x_i
        +
        F,
        \qquad
        i=1,\ldots,d,
        \label{eq:lorenz96}
    \end{align}
    with dimension $d=40$, forcing parameter $F=8$, and periodic boundary conditions
    $x_{-1}=x_{d-1}$, $x_0=x_d$, and $x_{d+1}=x_1$. The quadratic advection-like term produces local interactions among neighboring variables, allowing errors and observational information to propagate through the periodic state vector. This structure provides a more stringent test of ensemble-estimated cross-covariances than the Lorenz-63 system. The Lorenz-96 system is also integrated using \texttt{scipy.integrate.odeint} with adaptive time stepping.

\subsection{Observation Setup}
\label{subsection:obs_setup}

The numerical experiments are designed to evaluate the performance of the filtering methods under sparse observational settings, where only a subset of the model state variables is observed through the observation operators. In this setting, the EnSF-LR method propagates the state updates from the observed components to the unobserved components through the prior cross-covariance structure. The specific choice of observed variables can influence filter performance, since different observation networks constrain different dynamically important directions and provide different amounts of information about the unobserved state \citep{yang2006data,law2016filter}. Sensitivity to observation-network design has also been discussed in previous work on EnSF, where different spatial distributions and types of assimilated observations were shown to affect analysis accuracy \citep{xiong2026robustness}. In the present study, we therefore use fixed sparse observation configurations as controlled test cases, rather than attempting a comprehensive comparison of all possible observing networks.

\begin{itemize}
    \item \textbf{Observation Configuration:}
    For the Lorenz-63 model, only the $x$ component is observed, while the $y$ and $z$ components remain unobserved throughout the assimilation cycle. This choice provides a controlled sparse-observation setting in which a correction applied to a single observed variable must be mapped to the other two state variables either through the linear regression step or the model dynamics.
    Analogously, for the Lorenz-96 model, every fourth state variable is observed, corresponding to $25\%$ observational coverage of the 40-dimensional state vector. 

    For both the Lorenz-63 and Lorenz-96 models, observations are assimilated after 10 model integration steps, with the time step $\Delta t$ set to $0.02$.

    \item \textbf{Observation Operators:}
    We consider both linear and nonlinear observation operators:
    \begin{align}
        \text{Linear case:} \quad
        \mathbf{y}_t
        &=
        \mathcal{H}(\mathbf{x}_t)
        +
        \boldsymbol{\epsilon}_{l},
        \qquad
        \mathcal{H}(\mathbf{x}_t)
        =
        \mathbf{x}^{obs}_t,
        \label{eq:linear_obs_setup}
        \\
        \text{Nonlinear case:} \quad
        \mathbf{y}_t
        &=
        \mathcal{H}(\mathbf{x}_t)
        +
        \boldsymbol{\epsilon}_{nl},
        \qquad
        \mathcal{H}(\mathbf{x}_t)
        =
        \arctan(\mathbf{x}^{obs}_t),
        \label{eq:nonlinear_obs_setup}
    \end{align}
    where $\mathbf{x}^{obs}_t$ denotes the observed subset of the state vector. In the nonlinear case, the arctangent function is applied componentwise to the observed variables. The observation errors are Gaussian, with
    $\boldsymbol{\epsilon}_{l}\sim\mathcal{N}(\mathbf{0},\mathbf{R}_{l})$
    and
    $\boldsymbol{\epsilon}_{nl}\sim\mathcal{N}(\mathbf{0},\mathbf{R}_{nl})$.
\end{itemize}

\begin{table}[t]
\centering
\caption{Summary of the four experimental setups used to compare EnSF-LR against baseline methods.}
\label{tab:experiments}
\small 
\begin{tabular}{l|cc}
\toprule
\textbf{Operator} & \textbf{Lorenz-63} & \textbf{Lorenz-96} \\
\midrule
Linear & Exp\_L63: Linear obs. on Lorenz-63 & Exp\_L96: Linear obs. on Lorenz-96 \\
Nonlinear & Exp\_NL63: Nonlinear obs. on Lorenz-63 & Exp\_NL96: Nonlinear obs. on Lorenz-96 \\
\bottomrule
\end{tabular}
\end{table}

The observation error covariance is specified in terms of the observation error standard deviation. For the linear observation experiments, we use $\sigma_l=0.1$ for both Lorenz-63 and Lorenz-96, so that
$\mathbf{R}_l=\sigma_l^2\mathbf{I}_p$. This noise level is moderate relative to the typical dynamical range of the directly observed state variables in the linear-observation setting.
For the nonlinear observation experiments, smaller observation error standard deviations are used because the arctangent observation operator maps the state variables into a bounded range and compresses large-amplitude state variations. In the Lorenz-63 experiments, the observed $x$ variable typically spans approximately $[-20,20]$, while $\arctan(x)$ is bounded between $-\pi/2$ and $\pi/2$ and becomes relatively flat for large $|x|$. We therefore use $\sigma_{nl}=0.001$ for Lorenz-63, giving $\mathbf{R}_{nl}=\sigma_{nl}^2\mathbf{I}_p$. In the Lorenz-96 experiments, the observed variables have a smaller typical range, approximately $[-10,10]$, and we use $\sigma_{nl}=0.01$, again with $\mathbf{R}_{nl}=\sigma_{nl}^2\mathbf{I}_p$. Here, $\mathbf{I}_p$ denotes the identity matrix of dimension $p$, where $p$ is the number of observed variables.
\vspace{-1em}

\subsection{Evaluation Metric}

Filter accuracy is evaluated using the root-mean-square error (RMSE) between the analysis ensemble mean and the reference truth. Let $\bar{\mathbf{x}}_{t|t}$ denote the analysis ensemble mean at time $t$, and let $\mathbf{x}^{true}_t$ denote the corresponding reference state. The full-state RMSE is defined as
\begin{equation}
    \mathrm{RMSE}(t)
    =
    \frac{1}{\sqrt{d}}
    \left\|
    \bar{\mathbf{x}}_{t|t}
    -
    \mathbf{x}^{true}_t
    \right\|_2,
    \label{eq:rmse_full}
\end{equation}
where $d$ is the dimension of the full state vector and $\|\cdot\|_2$ denotes the Euclidean norm. When observed and unobserved components are analyzed separately, the same definition is applied to the corresponding subsets of the state vector.

For experiments averaged over multiple realizations, the time-dependent RMSE are first computed for each realization and then averaged across realizations.

\subsection{Ensemble Configuration}

\begin{itemize}
    \item \textbf{Reference Trajectory:}
    For each model, the reference truth is generated by integrating the governing equations from an initial state sampled from $\mathcal{N}(\mathbf{0},\mathbf{I})$. The model is integrated for $10^5$ numerical time steps with integration step size $\Delta t=0.01$. The first $5\times 10^4$ integration steps are discarded as spin-up. From the remaining trajectory, the reference truth is constructed by retaining every second integration step, resulting in an effective truth-output interval of $0.02$ model time units. From this saved trajectory, we select a continuous-in-time segment to serve as the reference truth for the assimilation experiment. The segment contains 300 saved reference states for Lorenz-63 and 600 saved reference states for Lorenz-96.
    \item \textbf{Initial Ensemble:}
    The initial ensemble is constructed from the spinned-up portion of the long model integration. Specifically, after integrating the model with numerical time step $\Delta t=0.01$, we randomly select $N_e$ time indices without replacement from the first $5\times 10^4$ integration steps. The model states at these selected times are used as the initial ensemble members. This time-lagged sampling strategy provides ensemble members that lie on, or near, the model attractor while retaining sufficient spread for ensemble-based estimation.
    \item \textbf{Ensemble Size:}
    We use $N_e=1000$ ensemble members for Lorenz-63 and $N_e=10,000$ ensemble members for Lorenz-96. These relatively large ensemble sizes are chosen deliberately to reduce the contribution of sampling error and to focus the experiments on the intrinsic approximation error of the filtering algorithms. Smaller ensemble sizes are important for practical applications, but they typically require additional algorithmic choices such as covariance localization and inflation. These choices can strongly affect filter performance and would make it more difficult to assess the baseline accuracy of the proposed EnSF-LR update. We therefore leave the systematic study of smaller ensembles and optimized localization/inflation strategies for future work.
    \item \textbf{EnSF Pseudo-time Discretization:}
    For the EnSF and EnSF-LR methods, the reverse SDE is integrated over the pseudo-time interval $\tau\in[1,0]$ using 500 uniform pseudo-time steps, with $\Delta\tau=1/500$. This pseudo-time discretization is used only for the score-based analysis update and is distinct from both the physical model time step and the assimilation interval.
\end{itemize}

\section{Results}
\label{section:results}

This section evaluates the data assimilation performance of the proposed Ensemble Score Filter with Linear Regression (EnSF-LR). The experiments assess whether the added regression step improves full-state estimation under sparse observations while preserving the score-based analysis update for observed state variables. We consider four experimental configurations using the Lorenz-63 and Lorenz-96 systems with both linear and nonlinear observation operators, as summarized in Table~\ref{tab:experiments}. EnSF-LR is compared with the original EnSF and with an EnKF baseline. The EnKF baseline represents the common update produced by the standard one-step EnKF and is equivalent to the two-step formulation, as shown both mathematically and numerically in Appendix~\ref{app:enkf_equivalence}. The results are organized by observation regime, beginning with linear observations and followed by nonlinear observations.

\subsection{Linear observations}

The linear observation experiments are motivated by a limitation observed in previous EnSF studies. In such idealized linear observation settings, EnKF can outperform the original EnSF because its covariance-based update propagates observational information across the full state. These experiments therefore test whether incorporating an EnKF-like regression mechanism into EnSF, through the proposed two-step EnSF-LR update, reduces this performance gap under sparse linear observations. Since the observation operator is linear in these experiments, the EnKF provides a strong covariance-based benchmark.

\begin{figure}[ht!]
\centering
\includegraphics[width=0.85\textwidth]{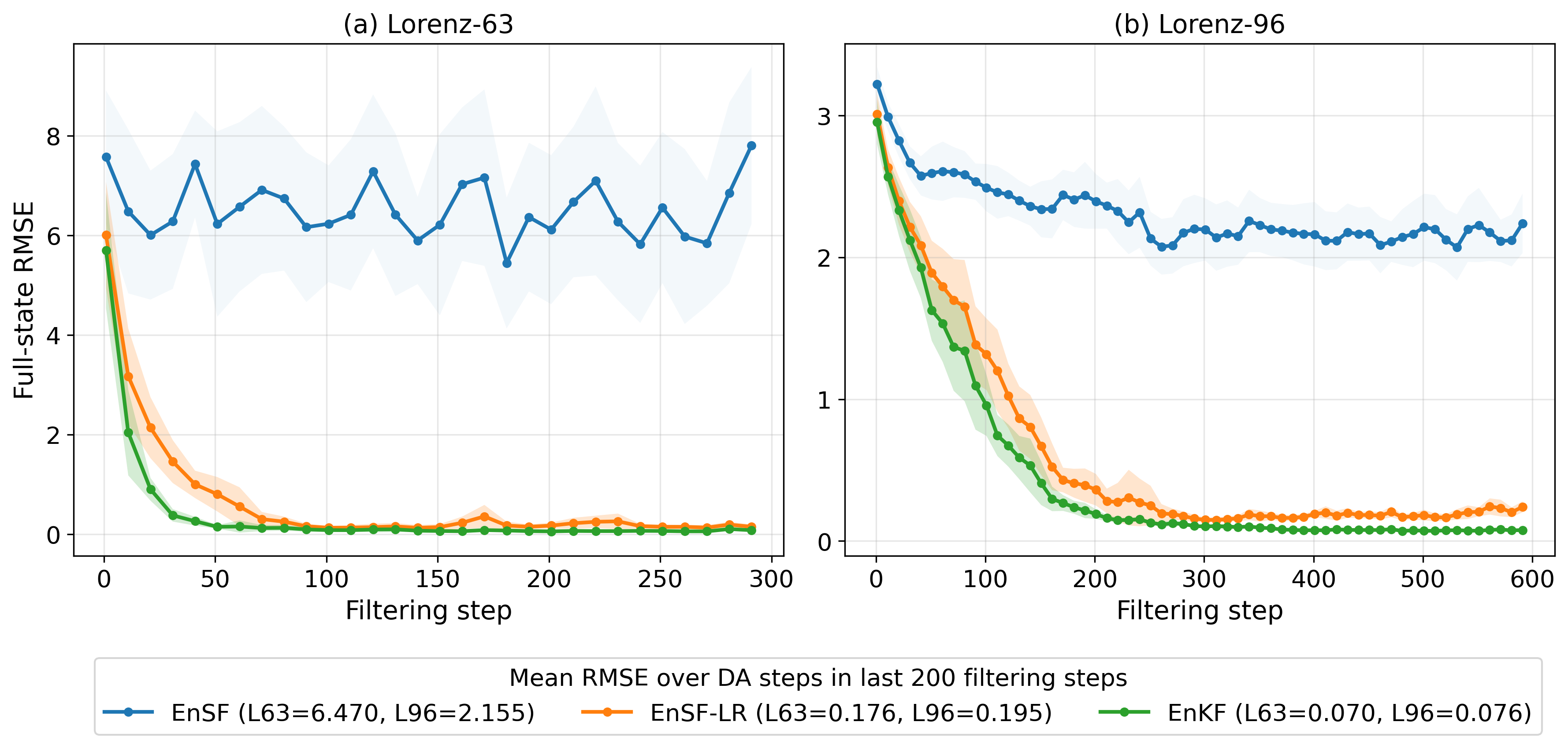}
\caption{
Analysis-time full-state RMSE for the linear-observation experiments in (a) Lorenz-63 and (b) Lorenz-96.
Curves show means over 20 independent realizations, and shaded bands indicate pointwise 95\% confidence intervals. Observations are assimilated every $k_{\mathrm{obs}}=10$ saved reference steps. Legend values are computed over assimilation times within the last 200 filtering steps.}
\label{fig:linear_rmse}
\end{figure}

Figure~\ref{fig:linear_rmse} shows the full-state RMSE for the Lorenz-63 and Lorenz-96 linear observation experiments. Across both models, EnSF-LR substantially reduces the RMSE compared with the original EnSF. This improvement is consistent with the purpose of the regression step. After the EnSF update corrects the directly observed variables, the resulting observed-state increments are mapped to the unobserved variables through the prior cross-covariance structure. The gain over the original EnSF is therefore mainly associated with the added propagation of analysis information to variables that are not directly observed.

The comparison with EnKF shows that EnSF-LR moves the original EnSF much closer to the covariance-based EnKF baseline, but does not completely reproduce the EnKF accuracy. In the Lorenz-63 experiment, EnSF-LR rapidly reduces the large errors of the original EnSF and reaches a low-error regime after the initial adjustment period. Its time-averaged RMSE over the last 200 filtering steps is much smaller than that of EnSF, although the EnKF baseline remains the most accurate. A similar pattern is observed in the Lorenz-96 experiment. EnSF-LR again produces a large reduction in RMSE relative to EnSF, while EnKF maintains the lowest mean RMSE. The remaining difference from EnKF is more apparent in Lorenz-96, where the higher-dimensional state places stronger demands on the ensemble-estimated cross-covariances used to regress observed-state increments onto unobserved variables.

Overall, these linear-observation results show that the regression step substantially reduces the performance gap between the original EnSF and covariance-based EnKF updates under sparse observations. This improvement is consistent across both the low-dimensional Lorenz-63 system and the higher-dimensional Lorenz-96 system. At the same time, the EnKF baseline remains more accurate in these linear settings, as expected from DA theory. We next consider nonlinear observation operators, where the score-based update in observation space is expected to provide a stronger advantage.

\subsection{Nonlinear observation}
Linear observation operators provide a useful baseline for assessing filter behavior, but many practical DA problems involve nonlinear relationships between model and observations variables. Such nonlinear observation operators can produce non-Gaussian likelihoods and pose additional challenges for filter stability and accuracy \citep{evensen2009data,hu2024pff,hu2025nongaussian_obs}. To examine the behavior of EnSF-LR in this more challenging setting, we repeat the Lorenz-63 and Lorenz-96 experiments using nonlinear observation operators, as defined in the Exp\_NL63 and Exp\_NL96 configurations in Table~\ref{tab:experiments}. These experiments are designed to assess whether the partitioned EnSF-LR update remains effective when the relationship between the state and observation spaces is nonlinear.

\subsubsection{Lorenz-63 model}
\label{subsection:4.b}

In the Exp\_NL63 experiment, the Lorenz-63 system is observed through the nonlinear arctangent observation operator. Only the $X$ component is observed, while the $Y$ and $Z$ components remain unobserved. Observations are assimilated intermittently, using the fixed observation interval described in Section~\ref{section:experimental_design}. This allows nonlinear forecast dynamics to act between analysis cycles and makes the prior distribution at each update more challenging for ensemble-based filters. The observation error standard deviation is reduced relative to the linear-observation experiment to account for the bounded range and nonlinear compression induced by the arctangent operator.

\begin{table}[t]
\centering
\caption{
95\% confidence intervals for the mean analysis-time full-state RMSE in the nonlinear-observation experiments.
For each realization, RMSE is averaged over assimilation times within the last 200 filtering steps.
Confidence intervals are computed across 20 independent realizations.
}
\label{tab:nonlinear_full_rmse_ci}
\begin{tabular}{lcc}
\toprule
Method & Lorenz-63 & Lorenz-96 \\
\midrule
EnSF    & $[6.085,\,6.575]$ & $[2.197,\,2.465]$ \\
EnSF-LR & $[2.555,\,2.937]$ & $[0.742,\,1.137]$ \\
EnKF    & $[3.836,\,4.561]$ & $[1.007,\,1.420]$ \\
\bottomrule
\end{tabular}
\end{table}

The full-state RMSE results in Fig.~\ref{fig:lorenz96_nonlinear_rmse} show that EnSF-LR provides the most accurate analysis among the three methods in the Lorenz-63 nonlinear-observation experiment.
Over assimilation times within the last 200 filtering steps, EnSF-LR achieves a mean full-state RMSE of $2.746$, compared with $6.330$ for the original EnSF and $4.198$ for the EnKF baseline. The corresponding 95\% confidence intervals are reported in Table~\ref{tab:nonlinear_full_rmse_ci}. Thus, EnSF-LR reduces the mean full-state RMSE by approximately 57\% relative to the original EnSF and by approximately 35\% relative to EnKF.

\begin{figure}[ht!]
\centering
\includegraphics[width=0.85\textwidth]{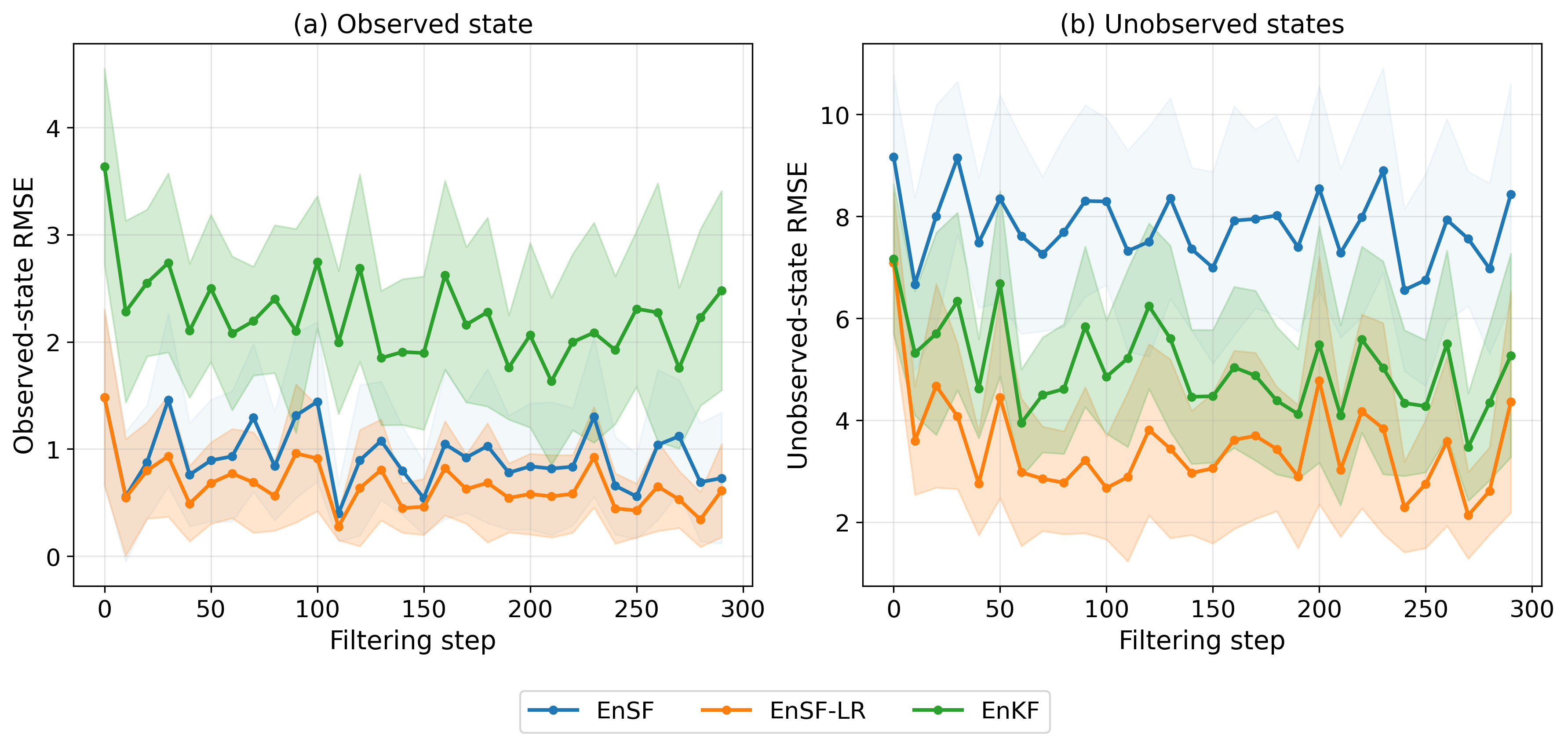}
\caption{
Observed- and unobserved-state RMSE for the Lorenz-63 nonlinear-observation experiment:
(a) observed state $X$ and (b) combined unobserved states $Y$ and $Z$.
Curves show mean RMSE evaluated only at analysis times and averaged over 20 independent realizations.
Shaded bands indicate pointwise 95\% confidence intervals for the mean across realizations.
Only the $X$ component is observed, while $Y$ and $Z$ are unobserved.
Observations are assimilated every $k_{\mathrm{obs}}=10$ saved reference steps.
}
\label{fig:lorenz63_nonlinear_rmse_obs_unobs}
\end{figure}

To identify where this full-state improvement comes from, Figure~\ref{fig:lorenz63_nonlinear_rmse_obs_unobs} separates the RMSE into observed- and
unobserved-state contributions. For the observed state $X$, EnSF-LR gives the smallest error among the three methods over most analysis times. The improvements over EnKF are consistent with the benefits of score-based DA updates under nonlinear observation operators. However, the difference between EnSF-LR and the original EnSF is more pronounced for the unobserved states $Y$ and $Z$, for which the original EnSF has the largest RMSE because its likelihood score correction affects only the observed $x$ component. 

\begin{figure}[ht!]
\centering
\includegraphics[width=0.85\textwidth]{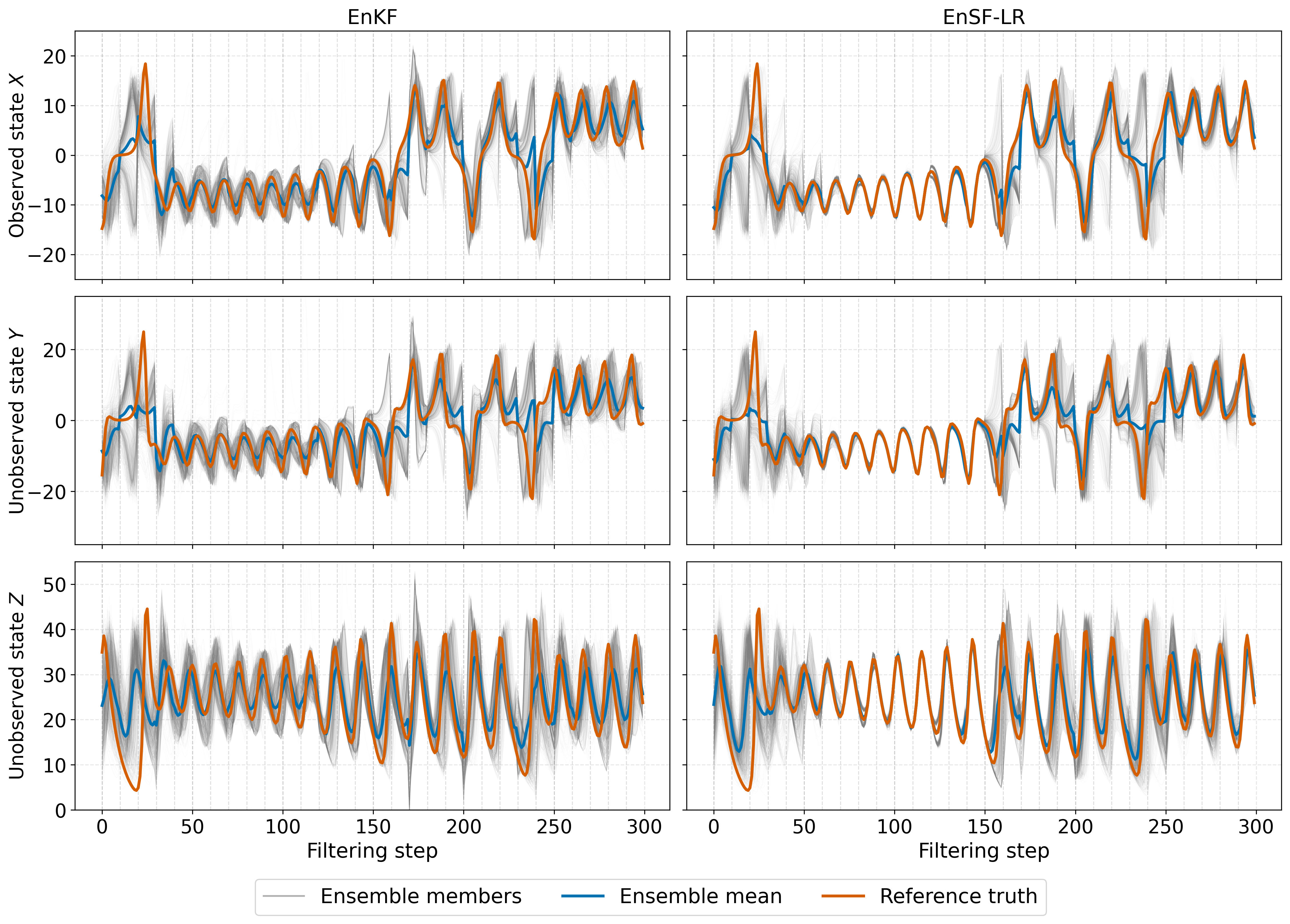}
\caption{
Representative ensemble trajectories for the Lorenz-63 nonlinear-observation experiment.
Rows show the $X$, $Y$, and $Z$ state variables, respectively. The left column shows EnKF and
the right column shows EnSF-LR. Gray curves denote ensemble members, the blue curve denotes
the ensemble mean, and the orange curve denotes the reference truth. Vertical dashed lines mark
assimilation times. Compared with EnKF, EnSF-LR keeps the ensemble trajectories more tightly
concentrated around the reference truth in this representative realization, particularly outside the
rapid transition interval identified in Fig.~\ref{fig:lorenz63_nonlinear_traj_3d}.
}

\label{fig:lorenz63_nonlinear_xyz_trajectories}
\end{figure}

The RMSE assesses the accuracy of the analysis mean, but it does not show how the ensemble evolves relative to the reference trajectory. To complement the RMSE diagnostics, we examine the ensemble trajectories for one representative realization in Figure Figure~\ref{fig:lorenz63_nonlinear_xyz_trajectories}. For the observed variable $X$, both filters follow the overall evolution of the truth after the spin-up portion of the displayed trajectory ($t \lesssim 50$). During the relatively smooth oscillatory intervals ($t \in [50,150]$ and $t \in [250,300]$), the EnSF-LR ensemble mean follows the truth more closely in both timing and amplitude, whereas the EnKF ensemble mean shows larger deviations from the truth. Similar behavior is seen for the unobserved variables $Y$ and $Z$. Since these components are not directly observed, their accuracy depends on how effectively information from the observed $X$ component is mapped to the unobserved state variables. The improved tracking of $Y$ and $Z$ by EnSF-LR is therefore consistent with the RMSE results and supports the role of the regression step in improving the unobserved-state analysis.

\begin{figure}[ht!]
\centering
\includegraphics[width=0.55\textwidth]{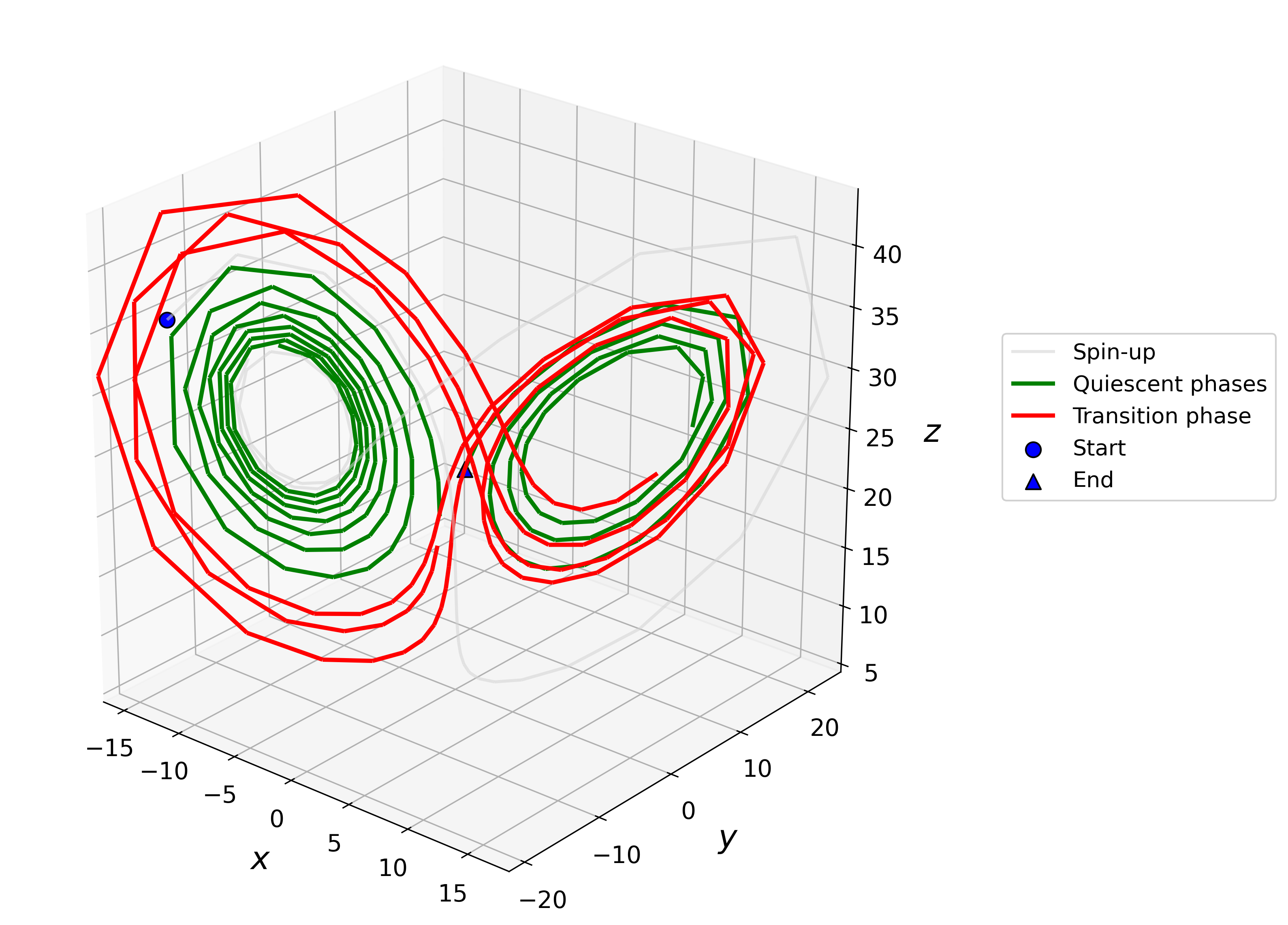}
\caption{
Three-dimensional reference trajectory of the Lorenz-63 system for the representative nonlinear-observation experiment.
The gray segment denotes the initial spin-up portion ($t \in [0,50]$), the green segments denote relatively smooth oscillatory phases ($t \in [50,150]$ and $t \in [250,300]$), and the red segment denotes the transition interval ($t \in [150,250]$) with rapid movement between the two lobes of the attractor.
The blue circle and blue triangle mark the start and end points of the displayed trajectory, respectively.
}
\label{fig:lorenz63_nonlinear_traj_3d}
\end{figure}

Both methods show reduced accuracy during the transition interval ($t \in [150,250]$), when the Lorenz-63 trajectory undergoes rapid movement between the two lobes of the attractor. These transitions are associated with stronger nonlinear error growth, making accurate state estimation more difficult between assimilation times. Figure~\ref{fig:lorenz63_nonlinear_traj_3d} illustrates these dynamical regimes along the three-dimensional trajectory of the truth.

\begin{figure}[ht!]
\centering
\includegraphics[width=0.7\textwidth]{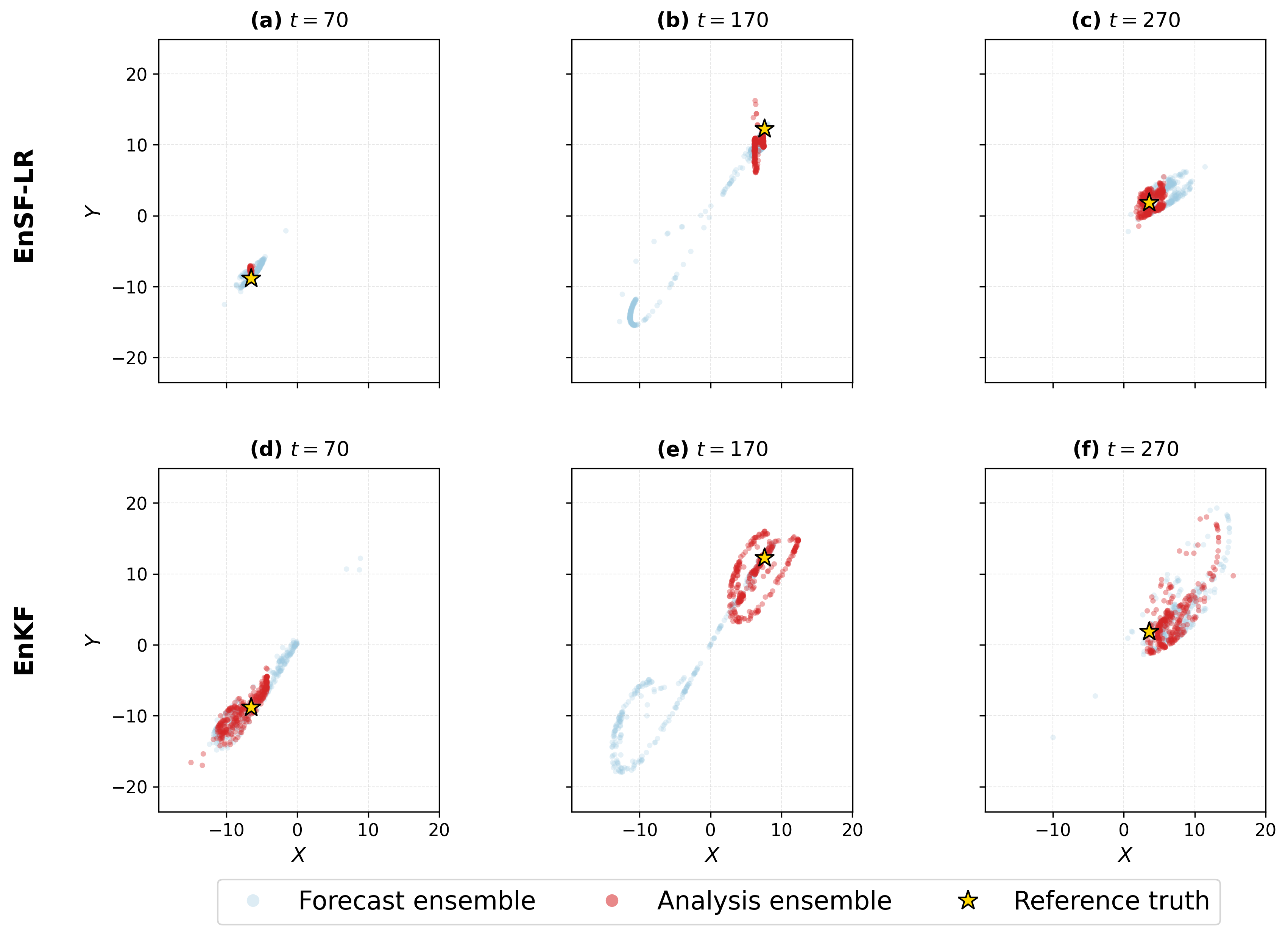}
\caption{
Forecast and analysis ensemble scatter plots in the $X$--$Y$ plane for the Lorenz-63 nonlinear observation experiment.
Columns correspond to representative analysis times $t=70$, $170$, and $270$.
The top row shows EnSF-LR and the bottom row shows EnKF.
Light blue points denote the forecast ensemble, red points denote the analysis ensemble, and the yellow star denotes the reference truth.
Only the $X$ component is observed, and in this projection, $Y$ is an unobserved state variable.
Results are shown for one representative realization with ensemble size 1000.
}
\label{fig:lorenz63_nonlinear_scatter}
\end{figure}

Additional phase-space diagnostics can be provided by ensemble scatter plots, which show how the analysis update modifies the forecast ensemble. Figure~\ref{fig:lorenz63_nonlinear_scatter} shows the forecast and analysis ensembles in the $X$--$Y$ plane at three representative times ($t=70$, $170$, and $270$). During the oscillatory phases ($t=70$ and $t=270$), both methods contract the forecast ensemble toward the truth after assimilating the observation. Compared with EnKF, EnSF-LR produces a more concentrated analysis ensemble around the true state in these snapshots, consistent with the lower RMSE and improved trajectory tracking discussed above. The EnKF forecast and analysis ensembles have larger spread, which is consistent with the larger RMSE of the EnKF ensemble mean in this representative case. The middle snapshot illustrates the difficulty of the transition regime. During this period, the forecast ensemble becomes strongly non-Gaussian and spreads along the curved structure of the Lorenz attractor. This makes the linear covariance-based EnKF update less effective, whereas the score-based EnSF-LR update gives an analysis ensemble that remains closer to the truth. Overall, these phase-space diagnostics provide qualitative support for the interpretation that EnSF-LR improves the analysis results under nonlinear observations, complementing the aggregate RMSE metrics.

\subsubsection{Lorenz-96 model}

The Lorenz-63 experiments provide a low-dimensional diagnostic of EnSF-LR under nonlinear observations. We next examine the Lorenz-96 system to assess the method in a higher-dimensional chaotic setting with sparse observations. It therefore provides a more demanding test of whether the proposed update can maintain accurate full-state estimates when observational information must be spread across a larger state vector.

\begin{figure}[ht!]
\centering
\includegraphics[width=0.85\textwidth]{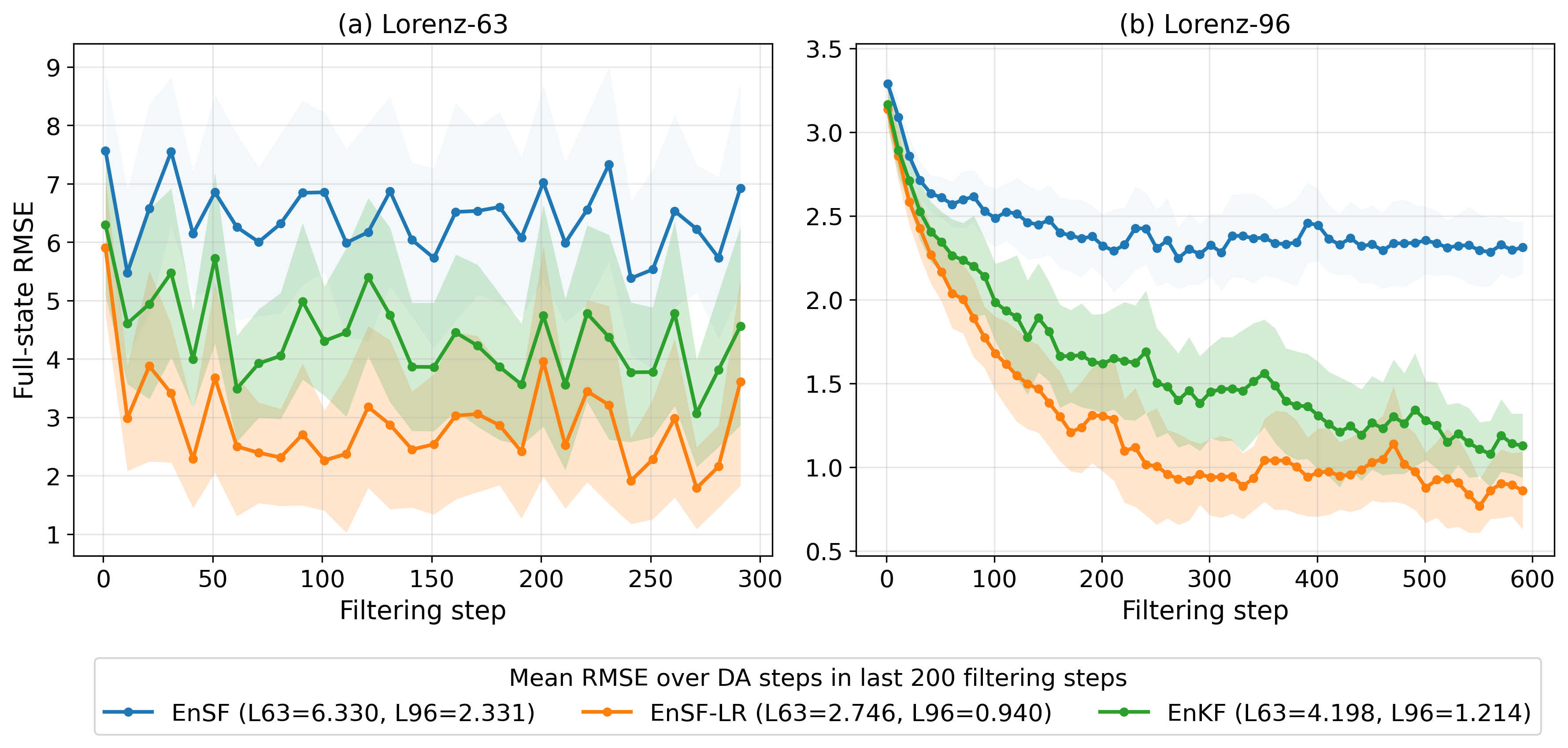}
\caption{
Analysis-time full-state RMSE for the nonlinear-observation experiments in (a) Lorenz-63 and (b) Lorenz-96.
Curves show the full-state analysis RMSE evaluated only at assimilation times and averaged over 20 independent realizations.
In the Lorenz-63 experiment, only the $X$ component is observed, while $Y$ and $Z$ are unobserved.
In the Lorenz-96 experiment, every fourth state variable is observed, corresponding to 25\% observational coverage.
Observations are assimilated every $k_{\mathrm{obs}}=10$ saved reference steps.
Mean RMSE values shown in the legend are computed over assimilation times within the last 200 filtering steps.
}
\label{fig:lorenz96_nonlinear_rmse}
\end{figure}

We repeat the nonlinear-observation experiments over 20 independent realizations. The aggregated full-state RMSE results are shown in Fig.~\ref{fig:lorenz96_nonlinear_rmse}, where each curve represents the analysis-time RMSE averaged across realizations and the shaded bands indicate pointwise 95\% confidence intervals for the mean. Mean RMSE values over assimilation times within the last 200 filtering steps are reported in the figure legend, and the corresponding 95\% confidence intervals are summarized in Table~\ref{tab:nonlinear_full_rmse_ci}. These results correspond to an approximately 60\% reduction relative to the original EnSF and an approximately 23\% reduction relative to the EnKF baseline.

The time series also shows that EnSF-LR maintains a lower RMSE than EnKF over most of the evaluation window, not only at isolated filtering steps. The original EnSF remains substantially less accurate, indicating that the score-based update in observation space alone is not sufficient when many state variables are unobserved. The EnSF-LR improvement therefore suggests that the higher-dimensional Lorenz-96 setting benefits from combining a nonlinear score-based update of the observed variables with regression-based propagation of information to the unobserved variables.

\begin{figure}[ht!]
\centering
\includegraphics[width=0.85\textwidth]{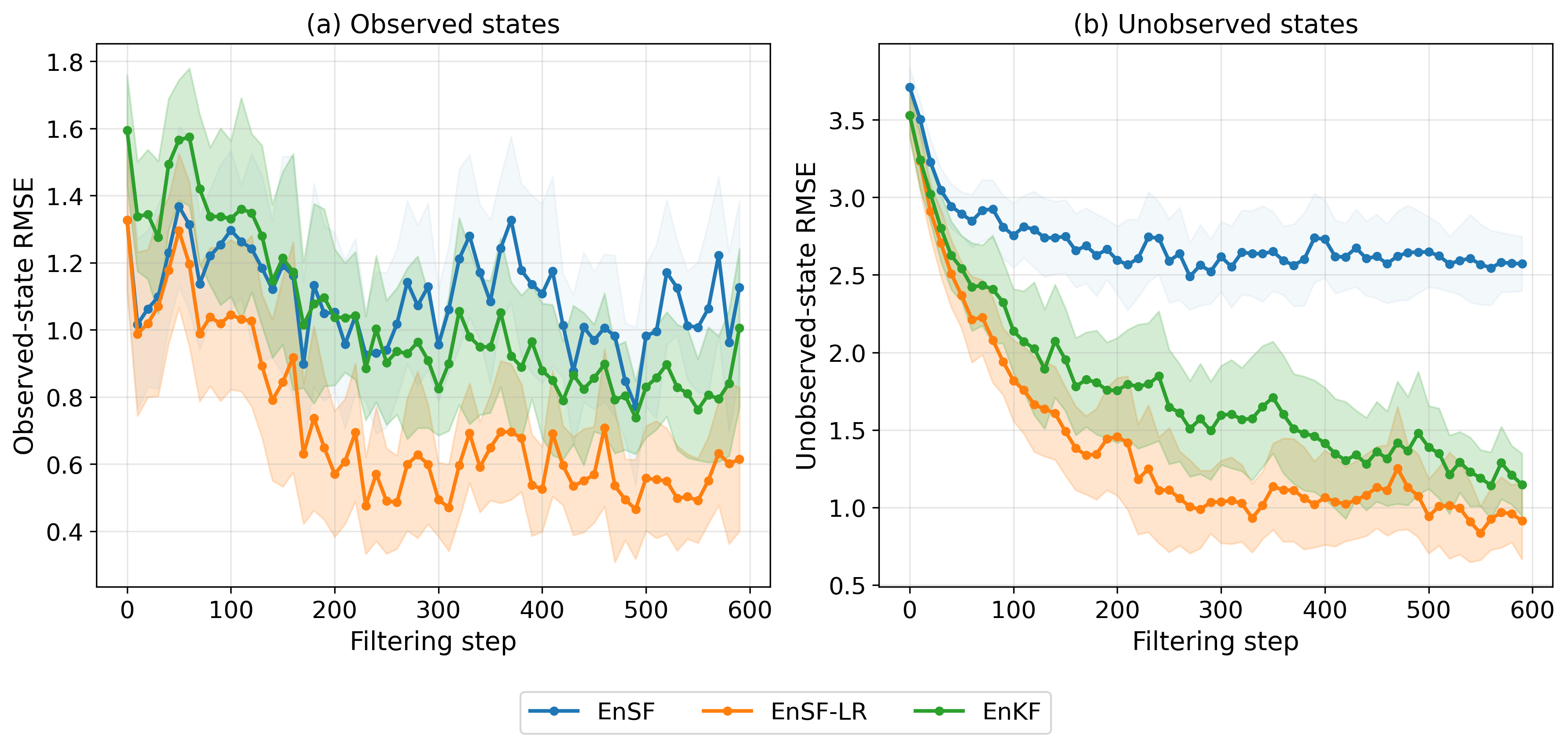}
\caption{
Observed- and unobserved-state RMSE for the Lorenz-96 nonlinear-observation experiment:
(a) observed state variables and (b) unobserved state variables.
Curves show mean RMSE evaluated only at analysis times and averaged over 20 independent realizations.
Shaded bands indicate pointwise 95\% confidence intervals for the mean across realizations.
Every fourth state variable is observed, corresponding to 25\% observational coverage.
Observations are assimilated every $k_{\mathrm{obs}}=10$ saved reference steps.
}
\label{fig:lorenz96_nonlinear_rmse_obs_unobs_avg}
\end{figure}

The decomposition into observed- and unobserved-state RMSE provides further insight into where the full-state improvement occurs. Figure~\ref{fig:lorenz96_nonlinear_rmse_obs_unobs_avg} shows the RMSE evaluated only at analysis times, separated into observed and unobserved state variables. For the observed variables, EnSF-LR consistently gives lower RMSE than the EnKF baseline, but the margin is relatively modest. The difference is more pronounced for the unobserved variables, where EnSF-LR produces a larger reduction in RMSE relative to both the original EnSF and EnKF. This indicates that the full-state improvement of EnSF-LR in the Lorenz-96 nonlinear experiment is driven primarily by better reconstruction of the unobserved variables, rather than by the observed variables alone.

\begin{figure}[ht!]
\centering
\includegraphics[width=0.85\textwidth]{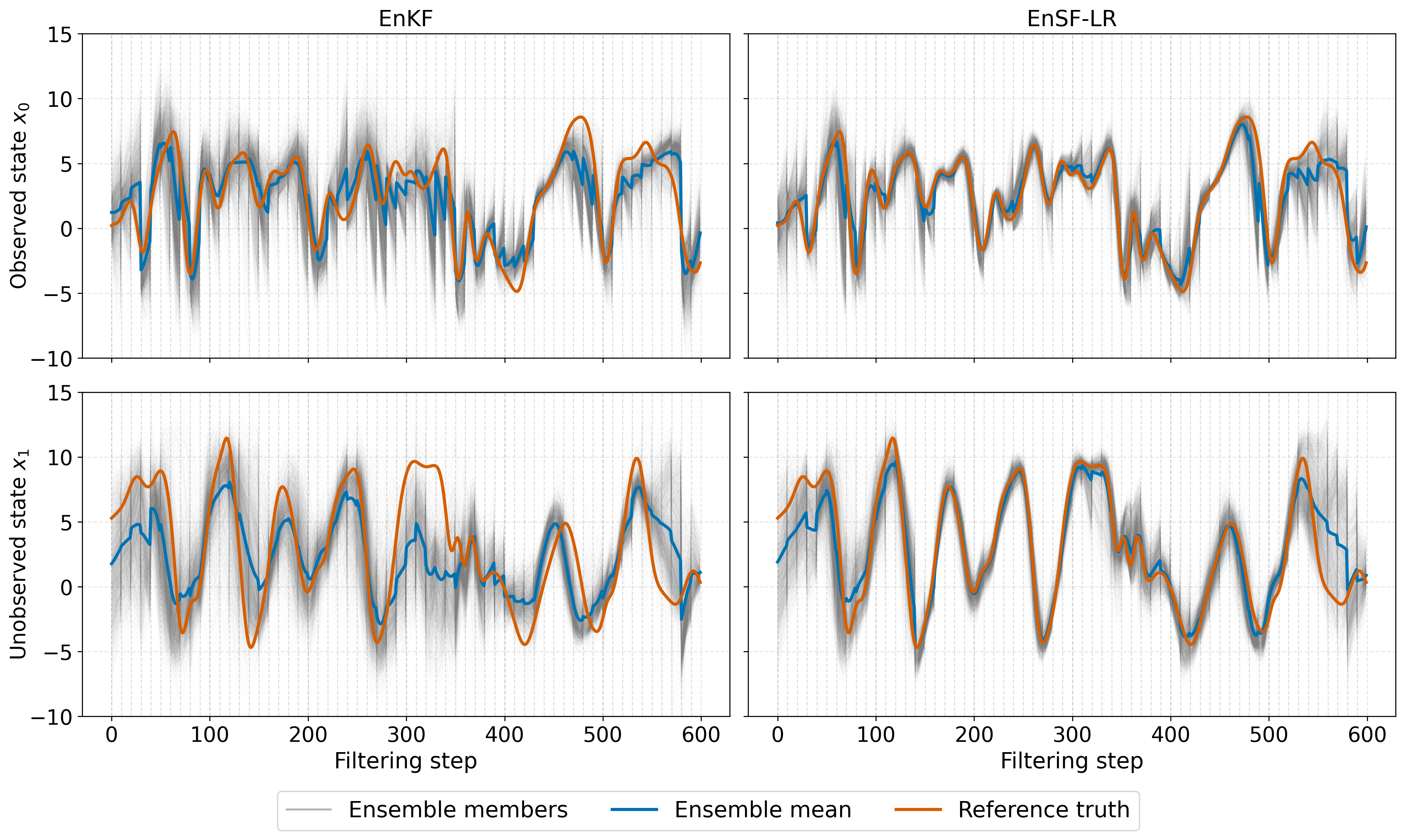}
\caption{
Representative ensemble trajectories for the Lorenz-96 nonlinear-observation experiment.
Rows show the directly observed state $x_0$ and the unobserved state $x_1$, respectively.
The left column shows EnKF and the right column shows EnSF-LR.
Gray curves denote a random subset of $2{,}000$ ensemble members drawn from the full $10{,}000$-member ensemble, the blue curve denotes the ensemble mean, and the orange curve denotes the reference truth.
Vertical dashed lines mark assimilation times.
Only every fourth state variable is observed, so $x_0$ is directly observed and $x_1$ is unobserved.
Results are shown for one representative realization.
}
\label{fig:lorenz96_nonlinear_x0_x1_trajectories}
\end{figure}

To complement the averaged RMSE diagnostics, we examine representative ensemble trajectories for one observed and one unobserved Lorenz-96 state variable in Figure~\ref{fig:lorenz96_nonlinear_x0_x1_trajectories}.
These trajectory plots provide a qualitative view of how the ensemble mean and ensemble spread evolve in time, and should be interpreted together with the 20-realization RMSE results in Figs.~\ref{fig:lorenz96_nonlinear_rmse} and~\ref{fig:lorenz96_nonlinear_rmse_obs_unobs_avg}.

For the Lorenz-96 trajectory comparison in Fig.~\ref{fig:lorenz96_nonlinear_x0_x1_trajectories}, both EnKF and EnSF-LR capture the main evolution of the observed variable $x_0$ in this representative realization. The difference is more visible for the unobserved variable $x_1$, where EnSF-LR tracks the reference truth more closely in several intervals. This is consistent with Fig.~\ref{fig:lorenz96_nonlinear_rmse_obs_unobs_avg}, where the improvement of EnSF-LR over EnKF is modest for the observed variables but more persistent for the unobserved variables.

This behavior differs from the Lorenz-63 trajectory comparison. In Lorenz-63, the main difficulty is associated with intermittent transitions between the two lobes of the attractor, leading to large oscillatory errors in the unobserved variables. In Lorenz-96, the selected variables do not show the same sharp regime-transition structure. Instead, the main challenge is maintaining accurate tracking in a higher-dimensional coupled system where information must be projected from observed to unobserved components. Thus, the Lorenz-96 trajectory result highlights sustained improvement in the unobserved state.

\begin{figure}[ht!]
\centering
\includegraphics[width=0.7\textwidth]{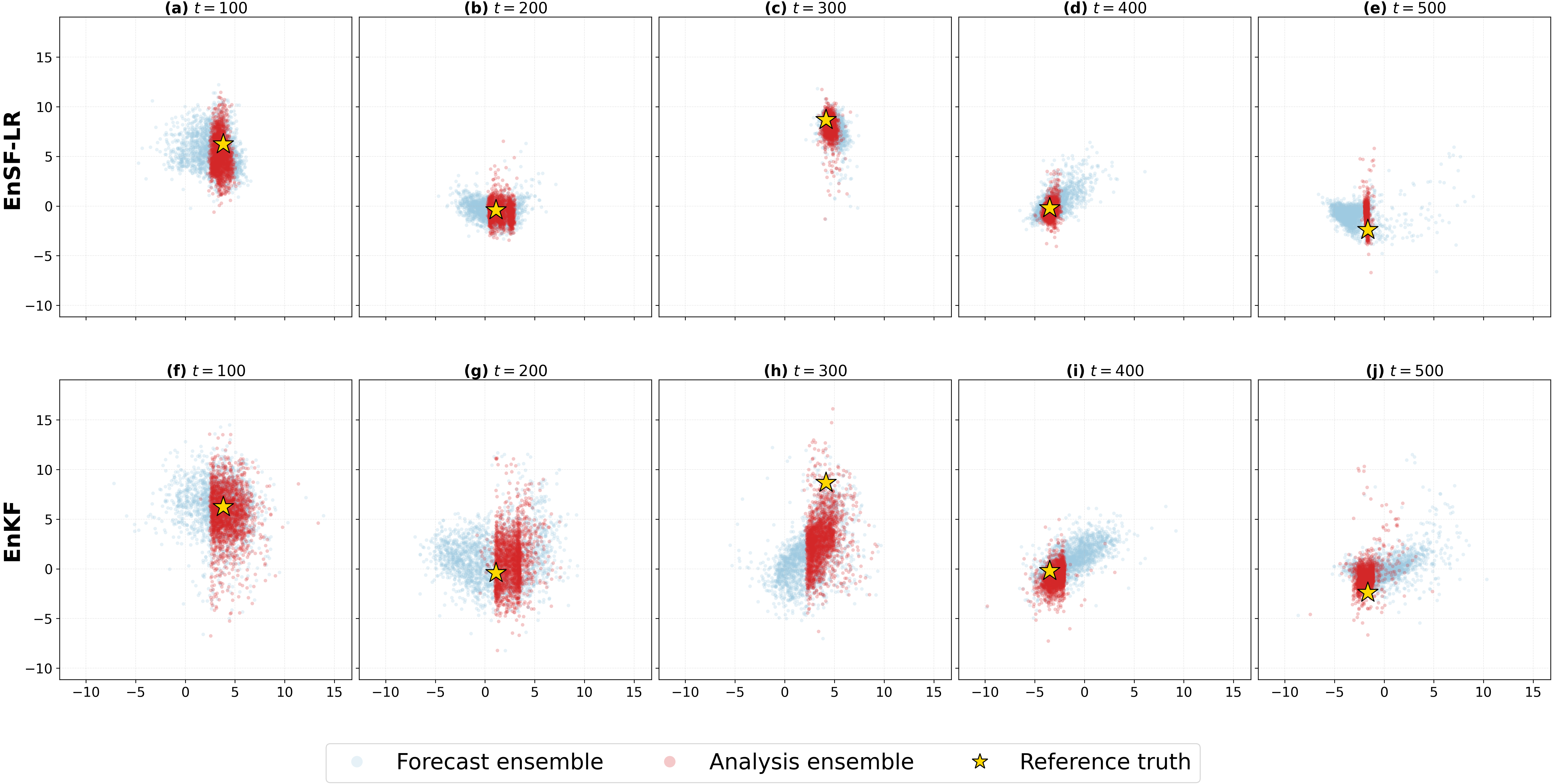}
\caption{
Forecast and analysis ensemble scatter plots in the $x_0$--$x_1$ plane for the Lorenz-96 nonlinear-observation experiment.
Columns correspond to representative analysis times $t=100$, $200$, $300$, $400$, and $500$.
The top row shows EnSF-LR and the bottom row shows EnKF.
Light blue points denote the forecast ensemble, red points denote the analysis ensemble, and the yellow star denotes the reference truth.
Every fourth state variable is observed. In this projection, $x_0$ is an observed state variable, whereas $x_1$ is unobserved.
Results are shown for one representative realization.
}
\label{fig:lorenz96_nonlinear_scatter}
\end{figure}

Finally, Fig.~\ref{fig:lorenz96_nonlinear_scatter} provides a phase-space view of the forecast and analysis ensembles in the $x_0$--$x_1$ plane for the same representative Lorenz-96 realization. Consistent with the RMSE and trajectory diagnostics, EnSF-LR tends to produce analysis ensembles that remain closer to the truth while retaining nonzero ensemble spread.
Compared with the Lorenz-63 scatter plots, the Lorenz-96 scatter plots show a less dramatic regime-transition structure in this two-dimensional projection. In Lorenz-63, especially during transitions between the two lobes of the attractor, the forecast ensemble can become strongly curved and visibly non-Gaussian. In the Lorenz-96 $x_0$--$x_1$ projection, the ensemble clouds are generally less curved, but the same qualitative pattern remains: EnKF analysis ensembles often exhibit larger displacements from the truth, whereas EnSF-LR yields a more centered analysis for both the observed variable and the neighboring unobserved variable.

\section{Conclusion}
\label{section:conclusion}

This study introduced the Ensemble Score Filter with Linear Regression (EnSF-LR), a two-step filtering method designed for data assimilation under sparse observations. The method addresses a limitation of the original EnSF in partially observed systems: the score-based likelihood update directly acts on observed state components, while unobserved components remain insufficiently corrected. EnSF-LR first applies the EnSF update to observed state variables and then maps the resulting increments to unobserved states through a linear regression step. Here, observed state variables are defined as the components of the model state that serve as inputs to the observation operator $\mathcal{H}(\cdot)$, while the remaining components are referred to as unobserved state variables.

The numerical experiments using the Lorenz-63 and Lorenz-96 systems demonstrate that this regression-based augmentation improves full-state estimation in sparse-observation settings. In the linear observation experiments, EnSF-LR produces RMSE values comparable to the EnKF while substantially improving upon the original EnSF.

The advantage of EnSF-LR becomes more pronounced in the nonlinear-observation experiments. For both Lorenz-63 and Lorenz-96, EnSF-LR reduces full-state RMSE relative to the original EnSF and achieves lower error than the EnKF in the tested nonlinear sparse-observation configurations. The observed- and unobserved-state RMSE diagnostics further show that the full-state improvement is obtained across both parts of the state vector, with the larger gains coming from the improved update of the unobserved state variables. Representative trajectory and phase-space diagnostics are consistent with this interpretation, showing improved tracking and more effective contraction of the forecast ensemble in the tested cases.

Overall, the results suggest that a hybrid DA strategy, which combines score-based  analysis in observation space with linear regression to unobserved state components provides a practical way to extend EnSF to partially observed nonlinear systems. The present study focuses on the low-dimensional Lorenz-63 and Lorenz-96 experiments with fixed but sparse observation networks. Future work will extend EnSF-LR to a serial ensemble DA framework compatible with the Data Assimilation Research Testbed (DART), a widely used community software for DA \citep{elgharamti2025dart,dartsoftware}. This extension would allow the proposed method to be tested with more realistic geophysical models and observation networks, while also enabling systematic studies of practical components such as localization and inflation.

\subsubsection{Acknowledgments}
The authors thank their colleagues and collaborators for helpful discussions and feedback during the development of this work. Z.X. acknowledges support from the Department of Mathematics at Florida State University. Additional funding and institutional support will be specified after peer review.

\subsubsection{Open-source code access}
The data used in this study were generated from numerical simulations of the Lorenz-63 and Lorenz-96 systems using the model equations, observation operators, and experimental settings described in the manuscript. No external observational datasets were used. The scripts used to generate the numerical results are available in the public GitHub repository: \url{https://github.com/ZixiangXiong/EnSF-LR}.

\clearpage
\appendix
\section{Equivalence of the One-Step and Two-Step Stochastic EnKF}
\label{app:enkf_equivalence}
\phantomsection

\setcounter{equation}{0}
\setcounter{figure}{0}
\renewcommand{\theHequation}{A.\arabic{equation}}
\renewcommand{\theHfigure}{A.\arabic{figure}}
\subsection{Algebraic equivalence}

This appendix shows that the one-step stochastic EnKF and the two-step
regression form of the stochastic EnKF give the same analysis update under the
covariance definitions used in this study. The equivalence holds when both
methods use the same forecast ensemble, the same perturbed observations, and
the same ensemble-based covariance matrices.

Let $\mathbf{x}_{t|t-1}^{(i)} \in \mathbb{R}^{d}$ denote the forecast state of
ensemble member $i$ at time $t$, and let
\begin{equation}
    \mathbf{y}_{t}^{f,(i)}
    =
    \mathcal{H}\left(\mathbf{x}_{t|t-1}^{(i)}\right)
    \in \mathbb{R}^{p}
\end{equation}
denote the corresponding forecast ensemble member in observation space. Let
$\bar{\mathbf{x}}^f_t$ and $\bar{\mathbf{y}}^f_t$ denote the forecast ensemble
means in state space and observation space, respectively. Define the scaled
forecast anomaly matrices
\begin{equation}
    \mathbf{X}
    =
    \frac{1}{\sqrt{N_e-1}}
    \left[
    \mathbf{x}_{t|t-1}^{(1)}-\bar{\mathbf{x}}^f_t,
    \ldots,
    \mathbf{x}_{t|t-1}^{(N_e)}-\bar{\mathbf{x}}^f_t
    \right],
\end{equation}
and
\begin{equation}
    \mathbf{Y}
    =
    \frac{1}{\sqrt{N_e-1}}
    \left[
    \mathbf{y}_{t}^{f,(1)}-\bar{\mathbf{y}}^f_t,
    \ldots,
    \mathbf{y}_{t}^{f,(N_e)}-\bar{\mathbf{y}}^f_t
    \right].
\end{equation}
The ensemble-based state--observation cross-covariance and observation-space
covariance are then
\begin{equation}
    \boldsymbol{\Sigma}_{xy}
    =
    \operatorname{Cov}(\mathbf{x}^f_t,\mathbf{y}^f_t)
    =
    \mathbf{X}\mathbf{Y}^{\top},
    \qquad
    \boldsymbol{\Sigma}_{yy}
    =
    \operatorname{Cov}(\mathbf{y}^f_t,\mathbf{y}^f_t)
    =
    \mathbf{Y}\mathbf{Y}^{\top}.
\end{equation}

In the stochastic EnKF, each ensemble member is updated using a perturbed
observation
\begin{equation}
    \mathbf{y}_{t}^{(i)}
    =
    \mathbf{y}_t
    +
    \boldsymbol{\epsilon}^{(i)},
    \qquad
    \boldsymbol{\epsilon}^{(i)}
    \sim
    \mathcal{N}(\mathbf{0},\mathbf{R}),
\end{equation}
where $\mathbf{R}$ is the observation error covariance matrix. The standard
one-step stochastic EnKF update is
\begin{equation}
    \mathbf{x}_{t|t}^{(i)}
    =
    \mathbf{x}_{t|t-1}^{(i)}
    +
    \boldsymbol{\Sigma}_{xy}
    \left(
    \boldsymbol{\Sigma}_{yy}
    +
    \mathbf{R}
    \right)^{-1}
    \left(
    \mathbf{y}_{t}^{(i)}
    -
    \mathbf{y}_{t}^{f,(i)}
    \right).
    \label{eq:app_onestep_enkf}
\end{equation}

The two-step form first updates the ensemble in observation space. The
observation-space analysis is
\begin{equation}
    \mathbf{y}_{t}^{a,(i)}
    =
    \mathbf{y}_{t}^{f,(i)}
    +
    \boldsymbol{\Sigma}_{yy}
    \left(
    \boldsymbol{\Sigma}_{yy}
    +
    \mathbf{R}
    \right)^{-1}
    \left(
    \mathbf{y}_{t}^{(i)}
    -
    \mathbf{y}_{t}^{f,(i)}
    \right).
    \label{eq:app_obs_update}
\end{equation}
Therefore, the observation-space increment is
\begin{equation}
    \Delta \mathbf{y}_{t}^{(i)}
    =
    \mathbf{y}_{t}^{a,(i)}
    -
    \mathbf{y}_{t}^{f,(i)}
    =
    \boldsymbol{\Sigma}_{yy}
    \left(
    \boldsymbol{\Sigma}_{yy}
    +
    \mathbf{R}
    \right)^{-1}
    \left(
    \mathbf{y}_{t}^{(i)}
    -
    \mathbf{y}_{t}^{f,(i)}
    \right).
    \label{eq:app_obs_increment}
\end{equation}

In the second step, this observation-space increment is mapped back to the model
state through linear regression:
\begin{equation}
    \Delta \mathbf{x}_{t}^{(i)}
    =
    \boldsymbol{\Sigma}_{xy}
    \boldsymbol{\Sigma}_{yy}^{-1}
    \Delta \mathbf{y}_{t}^{(i)}.
    \label{eq:app_regression}
\end{equation}
For notational simplicity, Eq.~\eqref{eq:app_regression} is written using
$\boldsymbol{\Sigma}_{yy}^{-1}$. If $\boldsymbol{\Sigma}_{yy}$ is rank deficient,
the same argument applies on the ensemble-based observation subspace using a
generalized inverse.

Substituting Eq.~\eqref{eq:app_obs_increment} into Eq.~\eqref{eq:app_regression}
gives
\begin{align}
    \Delta \mathbf{x}_{t}^{(i)}
    &=
    \boldsymbol{\Sigma}_{xy}
    \boldsymbol{\Sigma}_{yy}^{-1}
    \boldsymbol{\Sigma}_{yy}
    \left(
    \boldsymbol{\Sigma}_{yy}
    +
    \mathbf{R}
    \right)^{-1}
    \left(
    \mathbf{y}_{t}^{(i)}
    -
    \mathbf{y}_{t}^{f,(i)}
    \right) \nonumber \\
    &=
    \boldsymbol{\Sigma}_{xy}
    \left(
    \boldsymbol{\Sigma}_{yy}
    +
    \mathbf{R}
    \right)^{-1}
    \left(
    \mathbf{y}_{t}^{(i)}
    -
    \mathbf{y}_{t}^{f,(i)}
    \right).
    \label{eq:app_increment_equal}
\end{align}
Thus, the two-step state update becomes
\begin{equation}
    \mathbf{x}_{t|t}^{(i)}
    =
    \mathbf{x}_{t|t-1}^{(i)}
    +
    \boldsymbol{\Sigma}_{xy}
    \left(
    \boldsymbol{\Sigma}_{yy}
    +
    \mathbf{R}
    \right)^{-1}
    \left(
    \mathbf{y}_{t}^{(i)}
    -
    \mathbf{y}_{t}^{f,(i)}
    \right),
    \label{eq:app_twostep_equal}
\end{equation}
which is identical to the one-step EnKF update in
Eq.~\eqref{eq:app_onestep_enkf}.

This equivalence is not restricted to linear observation operators; it holds for any observation operator $\mathcal{H}$ when the same forecast (prior) ensembles and perturbed observations are used. It also remains valid with localization, inflation and observation perturbations, provided these are applied consistently in both formulations. Therefore, we report a single EnKF baseline in the main text and use the two-step form only as an implementation check and motivation for the regression step in EnSF-LR.

\begin{figure}[!htbp]
\centering
\includegraphics[width=\textwidth]{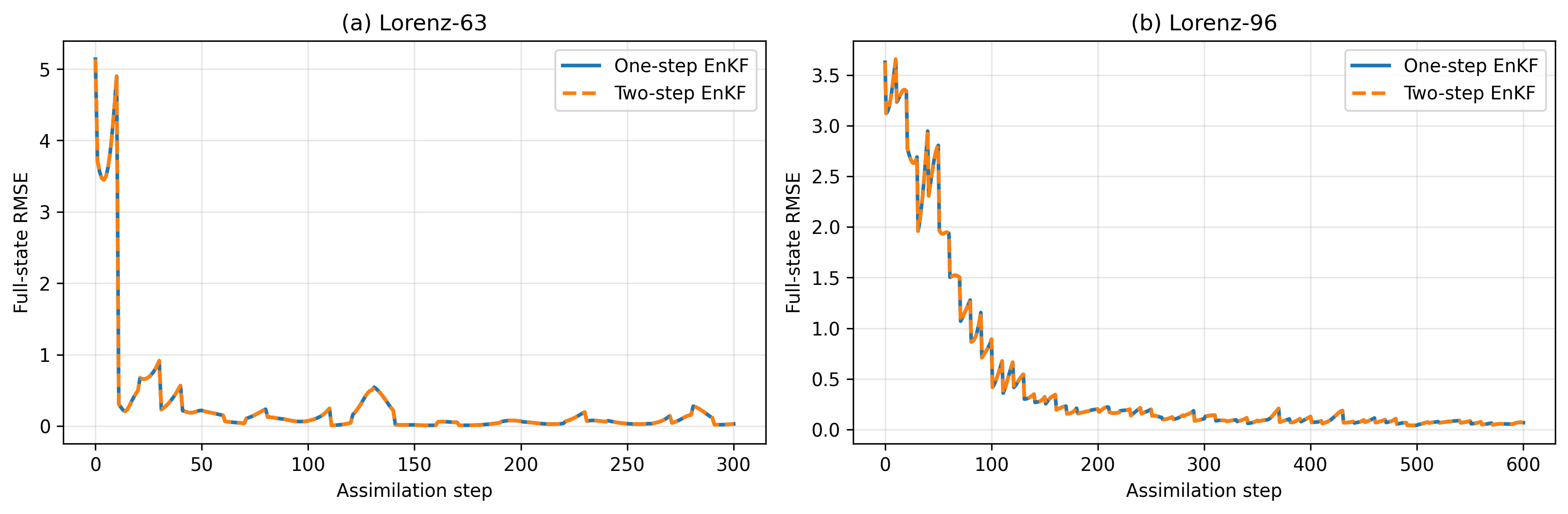}
\caption{
Numerical comparison of the one-step and two-step EnKF implementations for a single realization:
(a) Lorenz-63 with linear observations and (b) Lorenz-96 with linear observations. The overlapping
RMSE curves show that the two EnKF formulations give the same analysis results under the covariance
definitions used in this study, consistent with the algebraic equivalence derived above.
}
\label{fig:enkf_equivalence}
\end{figure}

As a numerical check, Fig.~\ref{fig:enkf_equivalence} compares the one-step
and two-step EnKF implementations for representative Lorenz-63 and Lorenz-96
linear-observation experiments. The overlapping RMSE curves confirm that the
two implementations produce the same analysis results in the settings considered
in this study.

\bibliographystyle{ametsoc2014}
\bibliography{references}
\end{document}